\documentclass{article}
\usepackage{arxiv} 
\usepackage[utf8]{inputenc}
\usepackage{color}
\usepackage{comment}
\usepackage{graphicx}
\usepackage{hyperref}

\newcommand{\citemiss}[1]{\textcolor{red}{(#1)}}

\title{Technological Factors Influencing Videoconferencing and Zoom Fatigue}
\author{ \href{https://orcid.org/0000-0002-9357-1763}{\includegraphics[scale=0.06]{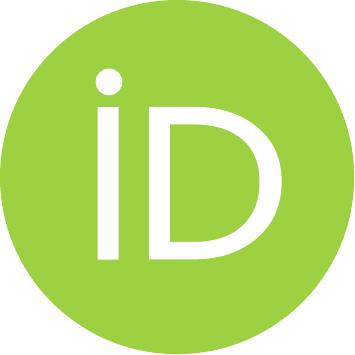}\hspace{1mm}Alexander Raake}\thanks{https://www.tu-ilmenau.de/en/mt-avt/} \\
	Institute of Media Technology\\
	Ilmenau University of Technology\\
	Ilmenau, Germany\\
	\texttt{alexander.raake@tu-ilmenau.de} \\
	\And
	\href{https://orcid.org/0000-0001-8929-4911}{\includegraphics[scale=0.06]{orcid.pdf}\hspace{1mm}Markus Fiedler} \\
	Department of Technology and Aesthetics\\
	Blekinge Institute of Technology\\ 
	Karlshamn, Sweden\\
	\texttt{markus.fiedler@bth.se} \\
    \And
	\href{https://orcid.org/0000-0001-9939-6671}{\includegraphics[scale=0.06]{orcid.pdf}\hspace{1mm}Katrin Schoenenberg} \\
	Department for Clinical Psychology\\
	and Psychotherapy\\ 
	Bergische Universität Wuppertal\\ 
	Wuppertal, Germany\\ 
	\texttt{schoenenberg@uni-wuppertal.de} \\
    \And
	\href{https://orcid.org/0000-0002-8752-3351}{\includegraphics[scale=0.06]{orcid.pdf}\hspace{1mm}Katrien De Moor} \\
	Department of Information Security and\\Communication Technology\\ 
	Norwegian University of Science and Technology\\ 
	Trondheim, Norway\\  
	\texttt{katrien.demoor@ntnu.no}
	\And
	\href{https://orcid.org/0000-0003-1299-4586}{\includegraphics[scale=0.06]{orcid.pdf}\hspace{1mm}Nicola D{\"o}ring} \\
	Institute of Media and Communication Science\\ 
	Technische Universität Ilmenau\\
	Ilmenau, Germany\\  
	\texttt{nicola.doering@tu-ilmenau.de}
}

\date{Version: \today}

\begin{document}

\maketitle

\begin{abstract}
The paper presents a conceptual, multidimensional approach to understand the technological factors that are assumed to or even have been proven to contribute to what has been coined as Zoom Fatigue (ZF) or more generally Videoconferencing Fatigue (VCF). With the advent of the Covid-19 pandemic, the usage of VC services has drastically increased, leading to more and more reports about the ZF or VCF phenomenon. The paper is motivated by the fact that some of the media outlets initially starting the debate on what Zoom fatigue is and how it can be avoided, as well as some of the scientific papers addressing the topic, contain assumptions that are rather hypothetical and insufficiently underpinned by scientific evidence. Most of these works are acknowledge the lacking evidence and partly suggest directions for future research. This paper intends to deepen the survey of VC-technology-related literature and to provide more existing evidence, where possible, while reviewing some of the already provided support or evidence for certain causal hypotheses. The technological factors dimension and its identified sub-dimensions presented in this paper are embedded within a more holistic four-dimensional conceptual factors model describing the causes for ZF or VCF. The paper describing this overall conceptual model is written by the same group of authors and currently under revision for an Open Access Journal publication. The present paper expands on the technological factors dimension descriptions provided in the overall model paper and provides more detailed analyzes and concepts associated with how VC technology may affect users' perception, cognitive load, interaction and communication, possibly leading to stress, exhaustion and fatigue. The paper currently is a living document which will be expanded further with regard to the evidence for or against the impact of certain technological factors.
\end{abstract}

\section{Introduction}

With the Covid-19 pandemic and the need for remote work and social encounters, the usage of VC technology has exploded during the past two years. 
After first seeing the term ``Zoom Fatigue'' (ZF) being mentioned in various media outlets, for example \cite{fosslien2020combat,kavanagh2021combat,jiang2020reason,Aili2021fatigue}, a number of scientific publications have appeared to more systematically analyze the effects, see, for example, \cite{bonanomi2021prevalence,kirk2020ll,riedl2021stress,williams2021working,bailenson2021nonverbal}.

The work by the authors of this paper in \cite{doering2022VCfatigueModel} describes a conceptual analysis of factors causing Videoconferencing Fatigue (VCF).
As most effects are not specific to the brand Zoom, the more general term VCF is used in the aforementioned and the present paper. 
Based on the analysis, a four-dimensional conceptual model of VCF is proposed in \cite{doering2022VCfatigueModel}.
The choice of these conceptual dimensions is based on an 8-step process that is described in detail in \cite{doering2022VCfatigueModel}, adapting the conceptual analysis approach by \cite{broder2019child}. 
It involves a scientific-paper- and media-outlet-based keyword and paper search, categorization and analysis, identification of dimensions and naming of components, and subsequently a multi-step refinement using discussions among authors, further paper searches about background concepts and empirical evidence, as well as feedback rounds and discussions within the authors' research groups. 
The resulting conceptual model comprising the four dimensions:
\begin{enumerate}
    \item Personal factors (individual and social factors)
    \item Organizational factors (temporal and context- or content-related factors)
    \item Technological Factors (presentation-related, communication-related, self-related and usability-related)
    \item Environmental factors (micro- and macro-environmental)
\end{enumerate}

Besides providing the novel, four-dimensional model, \cite{doering2022VCfatigueModel} investigates the evidence underlying specific hypotheses regarding factors that may cause fatigue. 
Accordingly, an individual evidence table is provided for each of the four dimensions, indicating whether a factor has been proven to cause fatigue or at least to lead to an increased mental load or effort.

The present paper is a somewhat refactored and considerably extended version of the technological factors dimension from \cite{doering2022VCfatigueModel} and initially serves as an Annex to that work. 
At this stage, it is a living document that will be expanded over time to possibly be proposed as a stand-alone publication.
The extended version was considered needed as the technological factors dimension is linked with a particularly large body of literature. 
Accordingly, in most previous media and scientific publications on ZF or VCF, the technological factor domain has been discussed with considerable emphasis \cite{bonanomi2021prevalence,kirk2020ll,riedl2021stress,williams2021working,bailenson2021nonverbal}.
This is because VC technology per se and accordingly a number of technology-related factors are what distinguishes VC from face-to-face (F2F) meetings or classical telephony. 
At the same time, the technology is what enables holding effective and efficient meetings with multiple participants over large distances. 
Yet, some technology-related factors are proven or very likely causes for videoconferencing fatigue, as will be discussed in this paper. 

This paper focuses on videoconferencing fatigue as a mental-psychological and possibly physical effect. 
Hence, the paper addresses technological characteristics of video conferences with regard to their effect on mental load or fatigue and categorizes the effects in terms of individual technology-related sub-dimensions. 
Relevant technological characteristics are outlined in the text as examples to illustrate the concrete technological causes of a given effect. 

The remainder of this paper is structured as follows: Section~\ref{sec:related} summarizes the findings from a few relevant scientific studies on technological factors associated with Zoom or VC fatigue. 
Section~\ref{sec:overview} describes the four sub-dimensions of the technological-factor dimension, following \cite{doering2022VCfatigueModel}.
The four sub-dimensions are addressed in one section each: Section~\ref{sec:presentation} deals with all effects that are related with auditory, visual and audiovisual perception.
In Section~\ref{sec:communication}, the role of technology for communication and interaction flow, the resulting cognitive load and possibly fatigue are discussed.
Section~\ref{sec:self} addresses technological factors that are related with oneself, such as the ease of producing intelligible speech sounds, being present on camera or sharing one's own room background during a VC meeting. 
The fourth sub-dimension on usability-related factors is discussed in Section~\ref{sec:usability}.
A set of conclusions are provided in Section~\ref{sec:conclusions}.

\section{Related work}
\label{sec:related}

Hacker et al. \cite{hacker2020virtually} carried out a comprehensive text mining analysis of about three million Twitter tweets from the beginning of the COVID-19 crisis in March 2020. 
Five affordances were identified for why users use VC technology, also referred to as ``action potentials'' in \cite{hacker2020virtually}: (A1) Communicating with social groups, (A2) engaging in shared social activities with family and friends, (A3) attending events, (A4) pursuing hobbies, and (A5) consuming non-recreational services, such as getting career advice or online appointments. 
Further, five constraints of VCs were identified that ``prevented or hindered people from accomplishing their goals'' \cite{hacker2020virtually}: (C1) Lacking features and competencies, (C2) having fear of being on camera, (C3) having to be always ``on'', (C4) exposing one's private living space, and (C5) lacking security. 
In the context of the present paper, it is noted that all of these constraints are related with specific properties of the VC technology used, as they are further analyzed in Section~3.3 of \cite{doering2022VCfatigueModel}, and further elaborated on in this paper.

Bailenson \cite{bailenson2021nonverbal} hypothesized a set of five technology-related, nonverbal mechanisms leading to cognitive overload, and thus Zoom fatigue: (F1) mirror anxiety, (F2) (being) physically trapped, (F3) hyper gaze, (F4) producing nonverbal cues, and (F5) interpreting nonverbal cues. 
Based on this set of possible nonverbal mechanisms for VC fatigue, Fauville et al. \cite{fauville2021zoom} introduced the Zoom Exhaustion \& Fatigue (ZEF) Scale. 
It addresses Bailenson's five mechanisms of fatigue (general, visual, emotional, social and motivational) with three items each, totaling to 15 items. 
The subsequent survey-based study reported in \cite{fauville2021nonverbal} contains a first indirect proof of the fatigue-impact of the five factors identified theoretically by \cite{bailenson2021nonverbal}, based on self-reports and with all data made available by the authors. 
Factors F1--F3 were assessed with multiple questions, while the remaining factors F4--F5 were addressed only by one question each. 
In their study, different effects of one factor mediating another were found. Moreover, a number of choices related to the study set-up may have influenced the results. More concretely, the scale creation process included crowdworkers from platforms such as Amazon Mechanical Turk. 
It is noted in this respect that the likely system-specific demographic sample implicitly involved may have led to a somewhat suboptimal selection of scale items, as specific types of VC usage may not have been well represented by the crowd. 
\cite{fauville2021nonverbal} have identified some further weak points of their study themselves, requiring further verification of the results. 
In this paper, the technology-related, nonverbal mechanisms investigated by \cite{bailenson2021nonverbal,fauville2021nonverbal} are analyzed in the context of the multidimensional factor approach proposed in \cite{doering2022VCfatigueModel} and discussed with regard to the underlying evidence. 
It is noted that the current version of this preprint will be expanded in this regard.

Based on a systematic review of the literature on ZF and VCF, Riedl  \cite{riedl2021stress} developed the following definition of Zoom Fatigue: ``Zoom fatigue (synonym: videoconference fatigue) is defined as somatic and cognitive exhaustion that is caused by the intensive and/or inappropriate use of videoconferencing tools, frequently accompanied by related symptoms such as tiredness, worry, anxiety, burnout, discomfort, and stress, as well as other bodily symptoms such as headaches.'' 
In reference to literature on the media naturalness theory \cite{kock2004psychobiological,kock2009information}, Riedl \cite{riedl2021stress} lists the features that characterize face-to-face (F2F) communication, the fulfillment of which determine the achieved level of naturalness of the communication: (1) the communicating individuals share the
same context, and they are able to see and hear each other, (2) they can quickly / in real-time exchange communicative stimuli, (3) the situation provides the ability to both convey and observe facial expressions, (4) to convey and observe body language, and (5) to convey and listen to speech. 
According to this line of thought, degradation in naturalness can lead to increased cognitive load and ultimately fatigue.
Based on his synthesis of media naturalness theory (cf. e.g., \cite{kock2005media}) and further literature, Riedl \cite{riedl2021stress} derived six technology-related root causes for VCF, three of which are deficiencies in comparison to F2F (1--3 below), and the other three (4--6) result from enrichments through software features that, according to \cite{riedl2021stress}, ultimately decrease naturalness and hence may increase fatigue: (1) Asynchrony of communication, (2) lack of body language, (3) lack of eye contact, (4) self-awareness, (5) unnatural interaction with multiple faces, and (6) multitasking opportunities.
It is noted that these factors bear similarities to the ones derived by \cite{bailenson2021nonverbal} and the constraints discussed in \cite{hacker2020virtually}, which were summarized above.
In \cite{riedl2021stress}, a conceptual model is proposed linking the different root causes to cognitive effort and stress.
Based on this, seven research hypotheses are established, the first six of which correspond to the root-causes, while the seventh hypothesizes the presence of a link between cognitive effort and stress.

The contribution of the present paper beyond the state-of-the art consists in a number of points.
First, it is embedded as part of the more holistic, four-dimensional conceptual model proposed in \cite{doering2022VCfatigueModel}. 
In the present paper it is reflected that previous research has more exclusively addressed the possible technological causes of VCF, which is considered in a deeper dive here than in \cite{doering2022VCfatigueModel}.
Second, the technological factors dimension itself consist of a set of sub-dimensions. Some of these correspond to aspects also discussed in other work such as \cite{bonanomi2021prevalence,kirk2020ll,riedl2021stress,williams2021working,bailenson2021nonverbal}. In this paper, however, they have been integrated in a more perception- and communication-centric paradigm.
Other sub-dimensions, in turn, have received less attention in previous literature, which are given more detailed consideration in this paper.
Third, both the accompanying, holistic framework paper \cite{doering2022VCfatigueModel} and the present, VC-technology-specific paper attempt to link the identified factor dimensions with the evidence available from the literature.
Here, a 4 $\times$ 3 categorization is employed in \cite{doering2022VCfatigueModel} and the present paper, considering the level of evidence available for a given factor\footnote{Four levels: (1) factors with proven impact on fatigue; (2) factors with link to effort or load that require further study (e.g., as there is no empirical evidence for the specific link between effort/load and fatigue); (3) factors that are likely linked to at least effort or load but without clear evidence and hence generally are for further study; (4) factors identified as irrelevant.} and the type of evidence data provided\footnote{Three levels: (1) Subjective, using scales and surveys as in, e.g., \cite{fauville2021nonverbal}; (2) objective, using metrics such as response times, success rates, or regarding load as in, e.g., \cite{tracy2020impact} for vocal load; (3) physiological, using indicators for effort, stress or fatigue, as e.g., in \cite{antons2012too,arndt2014low,arndt2016using}}.
It is noted, that especially regarding the evidence-related assessment, the present paper is a living work that will be updated up to a certain point, shall evidence data become available.
Besides these main contributions over the state-of-the-art, this paper contributes by including some complementary literature not previously considered in VC-fatigue-related research.

\section{Technological Factors}
\label{sec:overview}

As outlined above, the technological factors (TF) dimension \cite{doering2022VCfatigueModel} can be further subdivided into a total of four sub-dimensions. 
The rationale behind the technological factor's sub-dimensions is related with the perceptual and communication processes of persons involved in VC meetings:
\begin{enumerate}
    \item Presentation-related: Regarding the persons' auditory, visual and audiovisual perception per se, without specific focus on the communication-related role of perception.
    \item Communication-related: Reflecting the interaction and communication between persons and the role of signals and signs exchanged, the timing and the longer-term effects at the inter-personal level.
    \item Self-related: Addressing the processes that comprise self-related attentional aspects, self-perception and own actions as part of the communication. 
    \item Usability-related: Considering user-interaction with the system and associated effects due to lacking or hard-to use features and the associated technostress.
\end{enumerate}

Of the five constraints identified by Hacker et al. \cite{hacker2020virtually}, C1 (\textit{Lacking features and competencies}) and C5 (\textit{lacking security}) can be grouped in sub-dimension~4, C2 (\textit{having fear of being on camera}), C3 (\textit{having to be always ``on''}), and C4 (\textit{exposing one's private living space}) in sub-dimension 3. 

Of the five factors hypothesized by Bailenson \cite{bailenson2021nonverbal} and investigated empirically in \cite{fauville2021nonverbal}, the first two can be grouped into the self-related factors sub-dimension 3 (cf. Sec.~\ref{sec:self}), F1 (\textit{mirror anxiety}) and F2 (\textit{physically trapped}). The Factor F3 (\textit{hyper gaze}) is considered in the presentation-related factors sub-dimension 1 (Sec.~\ref{sec:presentation}). F4 (\textit{producing nonverbal cues}) and F5 (\textit{interpreting nonverbal cues}) are part of the communication-related factors sub-dimension 2 (Sec.~\ref{sec:communication}).

Of the six root-causes for VC fatigue identified by \cite{riedl2021stress}, (1) \textit{Asynchrony of communication}, (2) \textit{lack of body language}, (3) \textit{lack of eye contact} and (5) \textit{unnatural interaction with multiple faces} can be associated with the sub-dimension~2 of communication-related factors.
The root-cause (4) \textit{self-awareness} is comprised in the TF's sub-dimension~3 on self-related factors, and the root-cause (6) \textit{multitasking opportunities} is part of the TF sub-dimension~4 on usability-related factors (Sec.~\ref{sec:usability}). 


In the following, the four sub-dimensions will be discussed in more detail.

\section{Presentation-related Factors}
\label{sec:presentation}

The first sub-dimension contains all presentation-related factors. 
Here, all technological factors that characterize the one-way capture, processing and transmission of audio, video and audiovisual information are comprised. 
In this case, technological characteristics may increase the auditory or visual effort, or may lead to an increased effort for audiovisual integration of information.

\subsection{Visual Fatigue}
\label{sec:visual}

\emph{Visual effort} that may lead to fatigue: A link between visual fatigue and vergence effort for visual display units (VDUs) or visual display terminals (VDTs) has been reported in studies such as \cite{jaschinski1991}, \cite{tyrrell1990relation} and \cite{gur1994objective}. 
Earlier research on visual fatigue goes back to at least \cite{Donders1864} or \cite{BerensStark1932}, see also \cite{watten1994reinvention}.
Visual fatigue has been reported in various recent studies, and could for example be shown also for higher-resolution screens such as tablets \cite{kim2017visual}. 
As vergence rather than accommodation seems to drive fatigue, among other factors \cite{thomson1998eye}, the effect depends on the actual viewing distance from the VDU \cite{jaschinski1991}, \cite{gur1994objective}. 
From the literature it seems to be clear that any prolonged work on a PC or other near-viewing screen may cause some level of visual fatigue, and hence VC usage does not make a difference here. 
However, smaller distances to the screens are typical in VC than those from the other interlocutors in a face-to-face meeting - around 0.6\,m for close friends in face-to-face but much larger for acquaintances such as work colleagues, typically beyond 1.2\,m, according to the theory of \textit{proxemics} introduced by \cite{hall1966hidden}.
It can hence be assumed that eye strain may occur in case of screen usage during VC that is different from a natural face-to-face meeting (which can be concluded from work with slightly different focus such as \cite{hoffman2008vergence}).
As the aforementioned research has shown, prolonged work with computer screens fatigues ocular muscles, in contrast to natural viewing (cf., e.g., \cite{watten1994influence}).

Besides the factor of eye strain related to screen technology and viewing conditions, also the fact that VC does not enable the same natural ease of visual scene analysis in comparison to face-to-face, bears an influence. 
In face-to-face meetings, distance-dependent blurring of out-of-focus areas due to the depth-of-field of the human eye at closer focusing distances occurs, as well as binocular fusion vs. diplopia for in-focus vs. out-of-focus scene objects -- assuming participants are in the range of only few meters away from each other (in relation to the work on different social spaces introduced by \cite{hall1966hidden}). 
In VC, the recently introduced virtual backgrounds or background blurring may help to reduce the saliency of the real user backgrounds and help focus on the persons (extrapolating from the basic literature summarized, e.g., in \cite{hoffman2008vergence}).

Especially during the Covid-19 pandemic, participants primarily take part individually in front of their screens. 
Here, the scene layout differs considerably from the natural setting in a meeting room or social space like a kitchen or bar. 
With a screen, the switching and maintaining of visual attention is confined to a small space and field of view. 
Hence, mainly eye-movements are employed, instead of the larger movements of head and body involved in face-to-face communication, especially when orienting towards one of the multiple conversation partners. 
Note that in spite of the plausibility the authors did not find scientific evidence for an effect on fatigue for these specific aspects of visual scene-analysis. 

When analyzing visual scenes, a foreground-background separation is performed. In terms of technical factors, scene analysis of on-screen information may be affected by the chosen camera, lighting conditions, video resolution and coding, size of the participant’s video window on the screen, viewing distance and possible background effects such as virtual backgrounds or blurring. 
The fact that the ability of visual scene analysis and hence visual effort and ultimately visual fatigue depend on these different presentation-related factors has -- to the best of the authors' knowledge -- not been proven for all cases. 

However, there is strong evidence for this type of factors to play a role beyond the previously mentioned comparison of face-to-face versus screen-based viewing. 
For example, for older-generation screen technology with its lower resolutions, a distinct effect of resolution on eye strain and visual fatigue was identified. 
In their large-scale review of the so-called “computer vision syndrome”, \cite{Blehm2005} citing \cite{Ziefle1998} found that search reaction times and fixation durations during document viewing were significantly increased for the lower resolution, and that visual fatigue correlated with the search reaction times as well as eye movement parameters. 

For today’s PC and laptop displays used during VC sessions, screen resolution is generally higher than what was used in Ziefle’s studies \cite{Ziefle1998}. 
However, even for today’s videoconferencing technology, the displayed video resolution for some conversation partners may be considerably lower than the maximum given by the screen’s resolution. 
Also, sharpness or more generally visual quality may be further limited due to the camera technology and lighting situation during filming, or the video coding employed. 
Accordingly, the studies by \cite{Ziefle1998} and the review by \cite{Blehm2005} imply that visual effort and fatigue are induced when either the camera, the encoding or the display yield a lower visual resolution or general quality. 
Note that for the considerations on eye strain and visual fatigue it is assumed that the VC participants actually pay attention to what is shown on the screen, as attention mediates any visually induced fatigue effect, while fatigue in turn leads to loss of attention capability \cite{Boksem2005}. 
In an interactive communication scenario, attention to the other can mostly be assumed, if both sides have activated video and thus, the attention of the other person is tracked implicitly. Correspondingly, recommendations for reducing videoconferencing fatigue such as those by, for example, \cite{bennett2020examining} include the suggestion to turn off the camera as a means for temporally detaching, which will also reduce eyestrain.
The impact of mixed attention introduced by showing one's own picture along with that of the other interlocutors in today's VC solutions is addressed in the \emph{self-related factors} sub-dimension, Section~\ref{sec:self}.

\subsection{Related Theories}
There are a number of theories associated with the presentation-related technological factors sub-dimension, in particular:
\begin{itemize}
    \item Media Richness Theory (MRT)
          \cite{daft1983information,dennis1998testing}
    \item Media Naturalness Theory (MNT) \cite{kock2004psychobiological,kock2005media,kock2009information}
    \item Dual-Process Theory (DPT) 
          \cite{evans2013dual,kahneman2011thinking,ferran2008videoconferencing}
    \item Proxemics \cite{hall1966hidden}
\end{itemize}

\subsection{Auditory Fatigue}
\label{sec:auditory}

Besides visual effort, also increased \emph{listening effort} is a result of technical factors. 
It may be induced by the audio typical of VC and contribute to fatigue. 
Here, recent research for spatial versus non-spatial audio by \cite{fintor2021role} showed an increased listening effort for non-spatial audio even for only one single speaker attended to. 
Considering the role of spatial hearing in solving the so-called \textit{Cocktail Party} effect in a multiparty communication setting (cf.~e.g. \cite{bronkhorst2000cocktail}), and reduced speech intelligibility and hence listening effort in non-spatial audio settings, these results can safely be extrapolated to more cases with more than one source. 
It is noted that when listening to a VC via the computer’s built-in loudspeakers, the sound is listened to in a binaural manner. 
When listening to today’s VC solutions with headphones, in mostly all cases this involves diotic listening, as opposed to stereo or even binaural listening, thus without spatial audio cues that facilitate auditory scene analysis and corresponding speaker segmentation \cite{bregman1994auditory,bronkhorst2000cocktail}. The induced increased listening effort may be reduced by the fact that most VC meetings involve a sequential communication style with participants indicating that they wish to take the floor. 
In turn, with spatial audio, the more ecologically valid and improved usage of working memory \cite{baldis2001effects,skowronek2015assessment} and facilitated attention may lead to a better “focal assurance” \cite{baldis2001effects}, that is, better memorization of who said what in a VC-type meeting. 
These findings were made for spatial audio conferencing and not face-to-face communication with its natural binaural cues, and for that case, it is a logical assumption that listening effort and cognitive load will be further reduced in this case. 
Similarly, from the work by \cite{deng2019impoverished}, it can be deduced that the unrealistic auditory cues typical of today’s VC solutions, especially when listening with headphones, may reduce the ability of selective attention and hence increase cognitive load. 

Further, rather obvious audio-related factors that affect listening effort in terms of a Cocktail-Party type perspective are the employed audio signal level (``volume''); presence of background noise at the far or the near end (that is, the other parties’ environments or the own room); room reverberation in the own or other parties’s physical environment; the audio quality related with the microphones, loudspeakers or headphones used; degradations due to coding and possible packet losses during transmission, leading to impairments and possibly interruptions; or audio signal clipping due to echo cancellers and respective level-switching devices (for an overview see e.g.~\cite{raake2007speech}). 
All of these factors may affect audio quality. 
The link between audio quality and intelligibility has been discussed, for example, in \cite{raake2018telecommunications}. 
The relation between intelligibility and listening effort was discussed in various papers, such as e.g. \cite{krueger2017relation}. 
As reflected also by the aforementioned discussion on spatial versus non-spatial audio and cognitive load, also still intelligible speech but of somewhat reduced sound quality can be associated with an increased listening effort (see also recent work by \cite{WinnTeece2021}). 
In turn, it was shown recently in \cite{Rosset2021} that binaural hearing can improve comprehension, especially when participants in remote places need to wear face masks.

A series of studies conducted by \cite{antons2012too} and \cite{arndt2014low,arndt2016using} indicate that \emph{low-quality audio} as well as low-quality \emph{audiovisual} stimuli may lead to fatigue. 
From both shorter stimulus exposure and longer-sequence tests, EEG analysis points to an increase in workload and ultimately increased fatigue \cite{arndt2016using}. 
While the audiovisual stimuli used in \cite{arndt2016using} are not representative of VC content (a nature-related documentary was used with a narrator, instead of a conversation partner), the results are an indication that low-quality audiovisual stimuli may have a fatiguing effect. 
Interestingly, the subjective data on introspective fatigue judgments collected in parallel to the EEG data did not show any fatiguing effect, highlighting the usefulness of physiological data as a complement to questionnaires. 

\subsection{Related Theories}
There are a number of theories associated with the presentation-related technological factors sub-dimension, in particular:
\begin{itemize}
    \item Media Richness Theory (MRT)
          \cite{daft1983information,dennis1998testing}
    \item Media Naturalness Theory (MNT) \cite{kock2004psychobiological,kock2005media,kock2009information}
    \item Dual-Process Theory (DPT)
          \cite{evans2013dual,kahneman2011thinking,ferran2008videoconferencing}
    \item Auditory stream analysis
          \cite{bregman1994auditory}
\end{itemize}

\subsection{Audiovisual Fatigue}
\label{sec:audiovisual}

Audiovisual information integration in humans works seamlessly under normal, every-day situations. In a VC meeting, audio and video may be out of sync (see e.g.~\cite{berndtsson2018methods}). 
Detection tests indicate an asymmetry of the detection threshold, with shorter delays detected when audio is leading (between 15 to 50\,ms) and longer delays when video is leading (around 40 to 140\,ms), depending on the type of signal (see e.g., \cite{van2013rapid,Hollier1999,KohlrauschVandePar2005}). 
Possible explanations due to the difference in neural processing for the auditory and visual pathways have been proposed e.g., in \cite{KohlrauschVandePar2005}. 
To the best of the authors’ knowledge, no studies have been reported in the literature on the effect of audiovisual asynchrony on fatigue. 
Instead, the impact on or due to cognitive load has been addressed in some research. 
Work by \cite{eg2015audiovisual} and \cite{buchan2011cognitive} point only to a small effect of audiovisual asynchrony interaction with cognitive load. 
Instead, there seems to be the ability of the human perceptual system to adjust to certain levels of audiovisual asynchrony \cite{van2013rapid}.

Besides the understanding of the scene, there may also be an impact of the audiovisual presentation on the interpretation of the others or their actions. 
\cite{bailenson2021nonverbal} has hypothesized that unnaturally many staring faces might be present on one’s screen, with an intimidating effect. 
The questionnaire-based assessment conducted by \cite{fauville2021nonverbal} showed some support for this technology-induced factor. 
Yet, the used scale and intuitive logical assumption that this factor may be fatiguing may principally have led to some extent of bias in the respondents’ minds \cite{Orne2009}, a possible effect which is not further considered in \cite{fauville2021nonverbal}. 
A further presentation-related effect hypothesized by Bailenson  \cite{bailenson2021nonverbal} relates with Hall’s theory of \textit{proxemics} and social distance (cf.~\cite{hall1966hidden}).
Here, depending on the distance between camera and the distant person, on the camera field of view (i.e. whether wide-angle or rather long optics are being used), and on the distance of the viewer from the screen, the depicted face of the other may appear intimidatingly large on the listener’s screen. 
While the questionnaire-based study by Fauville et al. \cite{fauville2021nonverbal} seems to show some support for the validity of this factor for VCF, the practical relevance may be limited to persons with large screens and frequent one-to-one meetings. 
Some early evidence was provided in \cite{Ellis1992} that a smaller size of the face of an instructor shown on a screen can lead to better learning performance than a more intimidating closer shot. 
However, no direct relation to fatigue has been established so far, to the best of the authors’ knowledge. 
It is noted that a too loud audio signal may have a similar, undesirable impact on the listener. 
Here, however, the audio level is typically adjusted by the listener to a setting that is comfortable across all participants, or in case of a large difference between the levels of participants in a VC call, it may be indicated to a too-loud person to lower the microphone or speaking level, or to a too low person to get closer to the microphone, etc. (for further measures to improve sound quality by a listener, cf., e.g. \cite{raake2020binaural}). 

\subsection{Related Theories}
For this sub-dimension, some relevant underlying theories are:
\begin{itemize}
    \item Auditory scene analysis, Cocktail Party Effect  \cite{bregman1994auditory,cherry1953some,bronkhorst2000cocktail}
    \item Proxemics \cite{hall1966hidden}
\end{itemize}

\section{Communication-related Factors}
\label{sec:communication}

The second sub-dimension of the Technological Factors addresses effects disturbing interaction and communication that are caused by VC technology. 
According to \cite{egger2014quality}, interaction patterns can be defined as ``... a sequence of actions, references and reactions where each reference or reaction has a certain, ex-ante intended and ex-post recognisable, interrelation with preceding event(s) in terms of timing and content.''
Interactive human communication is characterized by different communication acts, that is, behavior interpreted by the other communication partner(s) and reacted upon accordingly (cf. e.g. \cite{Schonenberg2016PhD} based on \cite{petty1998attitude}).
When using VC technology, the fine-tuned interaction patterns that form the human communication flow may be altered or degraded.
As summarized for example by Skowronek \cite{skowronek2017PhD}, the two- or multiparty communication flow comprises different components. Important examples are: (i) turn taking, with verbal and/or non-verbal signs used to organize the change in the active speaker \cite{sacks1978simplest,knapp2013nonverbal}; (ii) back-channelling, in terms of signs conveyed to a currently active speaker by the other interlocutors, to signal attention, agreement or simply understanding \cite{o1993conversations}; or (iii) grounding \cite{clark1991grounding}, possibly involving multiple turns, with the aim to establish a common ground on a given part of the conversation.
All of the aforementioned components may be affected by VC technology deficiencies.

Three time scales can be distinguished for the interaction: 
\begin{enumerate}
    \item Short-term, involving nonverbal, vocal and non-vocal cues such as eye contact \cite{knapp2013nonverbal,schoenenberg2014interaction,bailenson2021nonverbal} and different acoustic or visual means for back-channeling  \cite{o1993conversations,knapp2013nonverbal}. 
    \item Mid-term, primarily addressing \emph{problems with turn taking}, cf. \cite{doering2022VCfatigueModel}. The concept of turns in conversations has been described, for example, by \cite{sacks1978simplest}. Effects of technical problems on turn taking and ultimately on VCF are discussed, e.g., in \cite{schoenenberg2014interaction,schoenenberg2014you, bailenson2021nonverbal,riva2021surviving}.
    \item Long-term, related with the communication process and possible, technology-induced misunderstandings of, or misconceptions about, the other(s) \cite{riva2021surviving,schoenenberg2014you,schoenenberg2014interaction}. In \cite{doering2022VCfatigueModel}, the impact on VC fatigue is referred to as \emph{problems with social bonding and impression formation}. Also included in effects of longer time scale are even more ``macroscopic'' aspects such as the ease of enabling meetings also at a distance, and holding meetings with reduced effort and travel. This may result in a positive effect with regard to stress and fatigue, that is, in a release from meeting fatigue (compare respective affordances identified by Hacker et al. \cite{hacker2020virtually}, A1, A2 and A3, see Sec.~\ref{sec:related}).
\end{enumerate}   

\subsection{Problems with Nonverbal Cues}
\label{sec:nonverbal}
Due to the importance of drawing meaning from head and eye movements for turn-taking and back-channeling \cite{sacks1978simplest,o1993conversations,knapp2013nonverbal}, and perceiving agreement and affection \cite{kleinke1986gaze}. For example Bailenson  \cite{bailenson2021nonverbal} pointed out that sending and receiving such hard-to-recognize non-verbal cues may cause additional mental load, introducing a corresponding risk for increased fatigue. This effect was supported with evidence from \cite{fauville2021nonverbal} and identified also by \cite{hacker2020virtually}. It is noted that the authors assume that counter-measures such as somewhat exaggerated signalling of the intent to take the turn (as indicated, for example, by the motion analysis in \cite{schoenenberg2014conversational}) may lead to increased exhaustion and hence VCF, too. 

Besides the lack of eye contact, the hypothesis has been raised that the overly high amount of apparent eye contact with many faces on screen may contribute to VC fatigue \cite{bailenson2021nonverbal}.
Similarly, Riedl \cite{riedl2021stress} has argued that the presence of multiple faces on screen causes stress, referring to aspects such as the ``eye contact effect'', see, e.g., \cite{senju2009eye}.
While the hypothesis appears compelling based on that literature, the evidence for this effect to play a major role for VCF appears questionable.
In particular, with an increased number of faces shown on the screen, the resulting small size of the faces as well as the typical encoding distortions, nonverbal cues such as eye contact are even harder to perceive, which likely intrinsically reduces the effect of the claimed ``multiple-faces'' hypothesis.
More empirical evidence is needed to substantiate the role of this possible factor.

\subsection{Problems with Turn Taking and Conversation Yiming}
\cite{powers2011effect} discusses the outstanding role of feedback delay for damaging interaction (as compared to reduced image quality and resolution), as it can interfere with the impression-formation process and increase cognitive load. This risk has also been pointed out by \cite{roberts2013identifying,schoenenberg2014you,schoenenberg2014interaction}. 
For example, \cite{schoenenberg2014interaction} quantified the degraded turn taking and resulting effects such as unintended interruptions. 
To re-organize the communication, additional effort is required, while some misunderstandings may not easily be recovered.
Accordingly, conversational games such as grounding may require effort and cause stress.
When noticing delay and attributing it to the system, the interlocutors may adjust their speaking and interaction behavior, yet hereby causing annoyance \cite{schoenenberg2014interaction}. 
While the impact of this factor on effort seems obvious, to the best of our knowledge, direct evidence for an impact on VC fatigue is lacking.

\subsection{Problems with Social Bonding and Impression Formation}
Instead of noticing effects such as delay as caused by the VC technology, they may be attributed to the communication partner, leading to interpersonal misjudgments \cite{powers2011effect}. 
In addition to causing momentary misunderstandings \cite{schoenenberg2014interaction}, the other may be considered as being less attentive, or the delay be attributed to personality traits such as extroversion, judging the other to be less friendly, less active, less cheerful, less self- efficient, less achievement-striving, and less self-disciplined \cite{schoenenberg2014you}.
Such misjudgments may also be caused by impaired audio and video, resulting in partly degraded verbal and especially non-verbal, affective cues, disrupting social bonding and impression formation \cite{roberts2013identifying,siegert2021case}. 
The result of the increased cognitive load may lead to VCF and \textit{Technostress} \cite{tarafdar2007understanding}. 
A further effect regarding impression formation is associated with Hall’s theory of proxemics and social distance \cite{hall1966hidden,bailenson2021nonverbal}.
The size of the other's face shown on one's own screen depends on the field of view of the other's camera and their distance from it, the own viewing distance from the screen, the screen size. According to \cite{bailenson2021nonverbal}, a large face on the screen may be perceived as intimidating. 
Considering average set-ups and on-screen VC window-sizes, the effect is likely rare. It is noted that \cite{fauville2021nonverbal} provided some survey-based support for this effect to occur. However, a formal empirical proof is not available.

\subsection{Effort Involved with Connection Problems}
Included in technological factors affecting VC fatigue are connection problems that may cause issues at any of the three aforementioned time scales, up to the need to re-establish the call. During a VC session, connection problems may lead to larger audio- or video-, or audiovisual impairments and interruptions. 
In such cases, intelligibility and visual information retrieval may temporarily be affected in a stronger manner, increasing cognitive load, cf. Section~\ref{sec:presentation}. 
This type of temporal effects that interrupt the conversation flow require an increased effort for information processing, both due to the involved increase in required perceptual resources at different levels, but also due to audiovisual attention switching from the interlocutors’ communication signals to the audiovisual-artefact stream. Moreover, some interaction with the VC system may be required to solve the problem, further disrupting the communication flow. 
There is a large body of research available on the cognitive load related to task switching, see e.g. \cite{liefooghe2008working,rubinstein2001executive}, which underlines the link to effort and ultimately fatigue. 
Technical problems during a VC session may be so severe, that one or several meeting participants need to completely leave the meeting and re-connect, or cannot continue to participate in the meeting, as a result of the technical problems. 
In this case, the entire communication is disrupted, at least for the participant that has the connection problems, but likely for all participants of the meeting. 
For other participants, the effect is mediated by the role of the persons affected in the meeting and the timing with regard to the meeting progress.
Such events cause annoyance for multiple participants, and \textit{Technostress} \cite{tarafdar2007understanding} for the person having connection issues, as well as those responsible for providing the meeting infrastructure (cf. Section~\ref{sec:usability}). 

\subsection{Other Effects} 
Discussions in the course of the preparation of this article within one of the author’s research groups pointed out the role of other, parallel communication channels such as Social Media or chat channels used during the VC. For example, a team that participates in a larger business meeting with further, external participants may use such a side-channel for communicating with a sub-group of participants in parallel to the larger meeting, possibly providing some release from VCF. There is no evidence supporting this consideration.

\subsection{Related Theories}
For the communication-related sub-dimension, relevant underlying theories are:
\begin{itemize}
    \item Media Richness Theory (MRT)
          \cite{daft1983information,dennis1998testing}
    \item Media Naturalness Theory (MNT) \cite{kock2004psychobiological,kock2005media,kock2009information}
    \item Dual-Process Theory (DPT) 
          \cite{evans2013dual,kahneman2011thinking,ferran2008videoconferencing}
    \item Turn taking, conversation analysis
          \cite{sacks1978simplest,hutchby2008conversation,sidnell2012handbook}
\end{itemize}

\section{Self-related Factors}
\label{sec:self}
The third sub-dimension contains self-related factors. Here, all factors are summarized that address signals that are produced by a given participant, such as their vocal statements during a VC meeting, as well as the perception of oneself during a VC meeting. Further included are aspects such as the enabled self-motion in front of the VC equipment.

\subsection{Problems with Being on Camera}
\label{sec:oncamera}

Of the self-related factors F1 and F2 hypothesized by \cite{bailenson2021nonverbal}, F1 represents the self-view, a.k.a. ``All Day Mirror''. 
This factor was shown to be mediated by the gender of the respective participant, with a higher relevance for female than for male participants. 
In an older study by \cite{wegge2006communication}, the effect of seeing oneself was found to be relevant for stress and negative emotions, especially in case that persons showed anxiety related to viewing and/or showing the own appearance.
In this regard, Hacker et al. \cite{hacker2020virtually} formulate ``Having fear of being on camera'' as constraint C2 of VC.

F2 addresses the limited movement range enabled by current VC camera set-ups, with participants basically being confined to the small field-of-view of the employed video camera. 
While these differ considerably across set-ups and some wide-angle lenses may still enable movement in front of the screen, in most cases the participants are seated and do not move around, whereas in face-to-face meetings they may be able to stand up and e.g., draw on a white board, etc. 
Limited mobility is also expressed in the VC-related constraint C3 described by \cite{hacker2020virtually}.
It is furthermore noted that \cite{riva2021surviving} imply that due to the neurophysiological linking of the location in which a certain event has occurred with episodic memory, the lack of a location in a VC session may lead to suboptimal memory usage, due to a ``loss of uniqueness of videoconferencing meetings''. 
According to \cite{riva2021surviving}, a further aspect of a personal experience affected by ``placelessness'' is the loss of a person’s sense of professional identity, due to the mixed roles.  

A further, connected self-view- and privacy-related factor is the aforementioned depiction of the real background in a person’s own environment and corresponding privacy concerns, see e.g. \cite{hacker2020virtually}, constraint C4 ``exposing one’s private living space''. 
Widely adopted measures to counterbalance obviously are muting the microphone to conceal actual sounds in one’s own space, turning off the camera (see e.g. \cite{bennett2020examining}), and choosing background blur or virtual backgrounds.
While turning off the camera may reduce both the privacy-related concerns as well as the ``All-day-mirror'' effect, social belongingness may be reduced (see also \cite{bennett2021videoconference}). 
To which extent virtual backgrounds or background blurring as offered by all major VC platforms really reduce VCF has, to the best of the authors’ knowledge, not been investigated to date. 
Recent work has, however, investigated the reasons why users select specific virtual backgrounds and the implication these may have on the perception of personality traits by the other participants \cite{hwang2021hide}. 
Here, the motivation varied depending on the ``perceived closeness with the other conversation partners'' and in terms of the specific individual intents when selecting a given background. 
The selected backgrounds were found to consistently have a ``muting'' effect, leading to neutralized ratings of personality traits.


\subsection{Problems with Producing Communication Xignals}
\label{sec:producing}
The authors of \cite{kristiansen2014study} found that school teachers talk with increased vocal load (i.e. a raised voice) in 61\% of time, mostly due to noise exposure. 
The authors were able to establish evidence for interrelations between raised exposure, raised speaking effort, reduced performance and cognitive fatigue after work. 
While no cases with VC-based teaching were included in that work, the parallels between real-life teaching and VC usage imply that a link between, e.g., higher effort in speaking and fatigue can be established also for VC. 
In the real-life data by \cite{kristiansen2014study}, the vocal effort depends also on the background noise. 
For VC-usage, it will further depend on the headphone's handling of one‘s own voice, talker echo, etc. Comparing face-to-face with remote audio-/visual communication,  \cite{tracy2020impact} identified an increase in vocal effort and produced speech sound pressure level for remote audio and remote audiovisual communication, as well as changes in other vocal parameters related with vocal effort and sound pressure.

The work presented in \cite{croes2019social} described several differences in nonverbal expressions between F2F and video-mediated communication. The latter implied more smiles, less facial touching and louder speech, while ``gaze aversion and a higher speech rate were found to influence social attraction'' \cite{croes2019social}. The authors highlight the importance of cue-rich forms of computer-mediated communication for social information processing. Combining these findings, we observe a causality chain from using VC technology via vocal effort to cognitive fatigue.

\subsection{Related Theories}
At this stage, only one related effect or theory is mentioned here, and the list will be expanded in future revisions of this paper:
Self-consciousness, Self-Attention, Assessment and Theory \cite{fenigstein1979self}.

\section{Usability-related Factors}
\label{sec:usability}

Last, but not least, videoconferencing implies that all participants must deal with the respective VC system.

As for any other ICT system, the \emph{offers of VC meet user expectations}. 
Two of the constraints identified by \cite{hacker2020virtually} from user complaints on Twitter address \emph{lacking features and competencies} (C1) as well as \emph{lacking security} (C5). 
The subjects in the study brought a set of work-arounds of missed-out features and set-up issues, and also report information overload and subsequent fatigue. 

Issues with mismatches between VC offers and expectations were furthermore taken up in discussions within the author’s research groups. 
In particular, people addressed the need to adapt to the capabilities offered by the VC, while creative spaces and gamification elements were missed out.
Any such concerns might trigger cognitive load and an emotional burden, leading to fatigue.

Handling a VC system’s diverse technological features can create \emph{Technostress} (\cite{tarafdar2007understanding}), which has been pointed out as a potential source of fatigue with negative consequences on productivity \cite{tarafdar2007understanding}, satisfaction \cite{Tarafdar2010} and performance \cite{Tarafdar2010}, \cite{Tarafdar2015}.
\cite{ayyagari35purvis} report a technostress-related field study involving 661 work professionals. 
Based on the person-environment fit model, they identified relationships between technology characteristics (such as usability, intrusiveness and dynamism) and stressors (such as work overload, work-home conflict etc.). 
\cite{pirkkalainen2019deliberate} points out the positive impact of IT control on user well-being because of reduced tensions and fatigue, referring to the results of \cite{galluch2015interrupting} that revealed a lower incidence of the stress hormone salivary amylase when being able to control the IT interruption. 

Technostress that might lead to \emph{techno-exhaustion}, expressed by \cite{weinert2020technostress} as ``effects of IT unreliability on end-user performance, techno-exhaustion, and physiological arousal, the direct influence of instrumental and emotional support on end-user performance, techno-exhaustion, and physiological arousal, and the results of a post hoc test to identify group differences in instrumental support and emotional support.'' 
Although \cite{weinert2020technostress} have not specifically addressed VC, we may deduce that technical difficulties and the lack of control lead to exhaustion and fatigue.

The degree to which operating a VC system is perceived as easy and convenient or stressful and fatiguing depends on the user’s characteristics such as their VC literacy (see \cite{doering2022VCfatigueModel}, Section 3.1) and on the \emph{system’s usability} and ease of use \cite{ayyagari35purvis, galluch2015interrupting, pirkkalainen2019deliberate}.

\subsection{Related Theories}
The main theory connected with this sub-dimension is that of Technostress, see, e.g., Tarafdar et al. \cite{tarafdar2007understanding} and Weinert et al. \cite{weinert2020technostress}.

\section{Conclusions}
\label{sec:conclusions}
The paper summarizes a number of technological factors that are associated with VCF or ZF.
These factors are grouped to factor dimensions related with specific processes associated with VC usage. 
The according technological factor dimension and its sub-dimensions form part of a more holistic, four-dimensional conceptual framework that also encompasses additional, non-technical factors \cite{doering2022VCfatigueModel}.
Besides the novel conceptualization, the paper expanded the previous collection of factors, constraints or root-causes \cite{hacker2020virtually,bailenson2021nonverbal,riedl2021stress} with additional factors.
Examples are contained in the \emph{presentation-related} factor dimension, which includes dedicated aspects around visual and auditory effort and also fatigue, or in the \emph{communication-related} factor dimension, which is related with typical communication processes.
Further, an in-depth analysis approach was adopted to identify whether specific factors, that is, sub-dimensions are proven to be factors that cause VC fatigue, or at least stress or load. 
In addition, the type of proof is addressed, i.e. whether empirical, subjective and questionnaire-based, performance-based or physiology-based assessment was undertaken.
From the analysis it can be concluded that only very few technical factors were scientifically proven to contribute to fatigue or stress. 
Of these, most have been assessed in surveys using questionnaires.
Here, it can be possible that some experimenter bias or effects due to demand characteristics may come into play \cite{Orne2009}, which may result in agreeing with statements about the impact of stress or fatigue when the answers appear logical or wanted in terms of experimental design.
Only for rather few sub-dimensions, results are available that are based on objective, either performance- or physiology-based, data.
It is noted that this specific analysis is one of the key aspects of this preprint paper, which will be expanded in future revisions.

\bibliographystyle{plain}
\bibliography{profRef}
\end{document}